\documentclass[twoside,twocolumn,english]{revtex4-1}
\usepackage[T1]{fontenc}
\usepackage[latin9]{inputenc}
\usepackage{geometry}
\usepackage{amsmath}
\usepackage{graphicx}

\makeatletter
\RequirePackage{colortbl, tabularx}
\@ifundefined{comment}{}% do nothing if the comment environment is not defined
  {% redefine the comment environment if it is defined
   
  }%

\makeatother

\usepackage{babel}
\begin{document}
\global\long\def\l{\lambda}%
\global\long\def\ints{\mathbb{Z}}%
\global\long\def\nat{\mathbb{N}}%
\global\long\def\re{\mathbb{R}}%
\global\long\def\com{\mathbb{C}}%
\global\long\def\dff{\triangleq}%
\global\long\def\df{\coloneqq}%
\global\long\def\del{\nabla}%
\global\long\def\cross{\times}%
\global\long\def\der#1#2{\frac{d#1}{d#2}}%
\global\long\def\bra#1{\left\langle #1\right|}%
\global\long\def\ket#1{\left|#1\right\rangle }%
\global\long\def\braket#1#2{\left\langle #1|#2\right\rangle }%
\global\long\def\ketbra#1#2{\left|#1\right\rangle \left\langle #2\right|}%
\global\long\def\paulix{\begin{pmatrix}0  &  1\\
 1  &  0 
\end{pmatrix}}%
\global\long\def\pauliy{\begin{pmatrix}0  &  -i\\
 i  &  0 
\end{pmatrix}}%
\global\long\def\pauliz{\begin{pmatrix}1  &  0\\
 0  &  -1 
\end{pmatrix}}%
\global\long\def\sinc{\mbox{sinc}}%
\global\long\def\ft{\mathcal{F}}%
\global\long\def\dg{\dagger}%
\global\long\def\bs#1{\boldsymbol{#1}}%
\global\long\def\norm#1{\left\Vert #1\right\Vert }%
\global\long\def\H{\mathcal{H}}%
\global\long\def\tens{\varotimes}%
\global\long\def\rationals{\mathbb{Q}}%
 
\global\long\def\tri{\triangle}%
\global\long\def\lap{\triangle}%
\global\long\def\e{\varepsilon}%
\global\long\def\broket#1#2#3{\bra{#1}#2\ket{#3}}%
\global\long\def\dv{\del\cdot}%
\global\long\def\eps{\epsilon}%
\global\long\def\rot{\vec{\del}\cross}%
\global\long\def\pd#1#2{\frac{\partial#1}{\partial#2}}%
\global\long\def\L{\mathcal{L}}%
\global\long\def\inf{\infty}%
\global\long\def\d{\delta}%
\global\long\def\I{\mathbb{I}}%
\global\long\def\D{\Delta}%
\global\long\def\r{\rho}%
\global\long\def\hb{\hbar}%
\global\long\def\s{\sigma}%
\global\long\def\t{\tau}%
\global\long\def\O{\Omega}%
\global\long\def\a{\alpha}%
\global\long\def\b{\beta}%
\global\long\def\th{\theta}%
\global\long\def\l{\lambda}%

\global\long\def\Z{\mathcal{Z}}%
\global\long\def\z{\zeta}%
\global\long\def\ord#1{\mathcal{O}\left(#1\right)}%
\global\long\def\ua{\uparrow}%
\global\long\def\da{\downarrow}%
 
\global\long\def\co#1{\left[#1\right)}%
\global\long\def\oc#1{\left(#1\right]}%
\global\long\def\tr{\mbox{tr}}%
\global\long\def\o{\omega}%
\global\long\def\nab{\del}%
\global\long\def\p{\psi}%
\global\long\def\pro{\propto}%
\global\long\def\vf{\varphi}%
\global\long\def\f{\phi}%
\global\long\def\mark#1#2{\underset{#2}{\underbrace{#1}}}%
\global\long\def\markup#1#2{\overset{#2}{\overbrace{#1}}}%
\global\long\def\ra{\rightarrow}%
\global\long\def\cd{\cdot}%
\global\long\def\v#1{\vec{#1}}%
\global\long\def\fd#1#2{\frac{\d#1}{\d#2}}%
\global\long\def\P{\Psi}%
\global\long\def\dem{\overset{\mbox{!}}{=}}%
\global\long\def\Lam{\Lambda}%
 
\global\long\def\m{\mu}%
\global\long\def\n{\nu}%

\global\long\def\ul#1{\underline{#1}}%
\global\long\def\at#1#2{\biggl|_{#1}^{#2}}%
\global\long\def\lra{\leftrightarrow}%
\global\long\def\var{\mbox{var}}%
\global\long\def\E{\mathcal{E}}%
\global\long\def\Op#1#2#3#4#5{#1_{#4#5}^{#2#3}}%
\global\long\def\up#1#2{\overset{#2}{#1}}%
\global\long\def\down#1#2{\underset{#2}{#1}}%
\global\long\def\lb{\biggl[}%
\global\long\def\rb{\biggl]}%
\global\long\def\RG{\mathfrak{R}_{b}}%
\global\long\def\g{\gamma}%
\global\long\def\Ra{\Rightarrow}%
\global\long\def\x{\xi}%
\global\long\def\c{\chi}%
\global\long\def\res{\mbox{Res}}%
\global\long\def\dif{\mathbf{d}}%
\global\long\def\dd{\mathbf{d}}%
\global\long\def\grad{\vec{\del}}%

\global\long\def\mat#1#2#3#4{\left(\begin{array}{cc}
#1 & #2\\
#3 & #4
\end{array}\right)}%
\global\long\def\col#1#2{\left(\begin{array}{c}
#1\\
#2
\end{array}\right)}%
\global\long\def\sl#1{\cancel{#1}}%
\global\long\def\row#1#2{\left(\begin{array}{cc}
#1 & ,#2\end{array}\right)}%
\global\long\def\roww#1#2#3{\left(\begin{array}{ccc}
#1 & ,#2 & ,#3\end{array}\right)}%
\global\long\def\rowww#1#2#3#4{\left(\begin{array}{cccc}
#1 & ,#2 & ,#3 & ,#4\end{array}\right)}%
\global\long\def\matt#1#2#3#4#5#6#7#8#9{\left(\begin{array}{ccc}
#1 & #2 & #3\\
#4 & #5 & #6\\
#7 & #8 & #9
\end{array}\right)}%
\global\long\def\su{\uparrow}%
\global\long\def\sd{\downarrow}%
\global\long\def\coll#1#2#3{\left(\begin{array}{c}
#1\\
#2\\
#3
\end{array}\right)}%
\global\long\def\h#1{\hat{#1}}%
\global\long\def\colll#1#2#3#4{\left(\begin{array}{c}
#1\\
#2\\
#3\\
#4
\end{array}\right)}%
\global\long\def\check{\checked}%
\global\long\def\v#1{\vec{#1}}%
\global\long\def\S{\Sigma}%
\global\long\def\F{\Phi}%
\global\long\def\M{\mathcal{M}}%
\global\long\def\G{\Gamma}%
\global\long\def\im{\mbox{Im}}%
\global\long\def\til#1{\tilde{#1}}%
\global\long\def\kb{k_{B}}%
\global\long\def\k{\kappa}%
\global\long\def\ph{\phi}%
\global\long\def\el{\ell}%
\global\long\def\en{\mathcal{N}}%
\global\long\def\asy{\cong}%
\global\long\def\sbl{\biggl[}%
\global\long\def\sbr{\biggl]}%
\global\long\def\cbl{\biggl\{}%
\global\long\def\cbr{\biggl\}}%
\global\long\def\hg#1#2{\mbox{ }_{#1}F_{#2}}%
\global\long\def\J{\mathcal{J}}%
\global\long\def\diag#1{\mbox{diag}\left[#1\right]}%
\global\long\def\sign#1{\mbox{sgn}\left[#1\right]}%
\global\long\def\T{\th}%
\global\long\def\rp{\reals^{+}}%

\title{Origin of universality in the onset of superdiffusion in L\'evy walks}
\author{Asaf Miron}
\address{Department of Physics of Complex Systems, Weizmann Institute of Science,
Rehovot 7610001, Israel}
\begin{abstract}
Superdiffusion arises when complicated, correlated and noisy motion
at the microscopic scale conspires to yield peculiar dynamics at the
macroscopic scale. It ubiquitously appears in a variety of scenarios,
spanning a broad range of scientific disciplines. The approach of
superdiffusive systems towards their long-time, asymptotic behavior
was recently studied using the L\'evy walk of order $1<\b<2$, revealing
a universal transition at the critical $\b_{c}=3/2$. Here, we investigate
the origin of this transition and identify two crucial ingredients:
a finite velocity which couples the walker's position to time and
a corresponding transition in the fluctuations of the number of walks
$n$ completed by the walker at time $t$.
\end{abstract}
\maketitle
\textit{Introduction} - Diffusion effectively models the dynamics
of many physical systems. Its hallmark property, a linear increase
of the mean-square displacement (MSD) with time, famously describes
the stagnant motion of a grain of pollen tumbling about in a glass
of water \citep{brown1828xxvii}. Yet there is an ever-growing list
of ``superdiffusive'' phenomena that fall well outside the paradigm
of simple diffusion, in which perturbations propagate faster than
diffusion. Notable examples include the dynamics of turbulent systems
\citep{shlesinger1987levy}, spreading of perturbations and associated
1D anomalous transport \citep{cipriani2005anomalous,zaburdaev2011perturbation,liu2012anomalous,dhar2013exact,cividini2017temperature,PhysRevE.100.012106},
tagged particle dynamics in disordered media \citep{levitz1997knudsen,brockmann2003levy},
evolution of trapped ions and atoms in optical lattices \citep{marksteiner1996anomalous,katori1997anomalous,sagi2012observation},
and even the behavior exhibited by living matter \citep{PhysRevLett.65.2201,PhysRevE.47.4514,upadhyaya2001anomalous,reynolds2018current,rhee2011levy,raichlen2014evidence}.

The L\'evy walk is a canonical model for superdiffusion. In 1D, it
describes a \textquotedblleft walker\textquotedblright{} evolving
in a series of independent ``walks''. At the start of each walk,
the walker randomly draws a \textquotedblleft walk-time'' $\t$ and
a direction $\pm1$ along which it moves with velocity $\pm v$ for
the duration of the walk. Superdiffusion arises when the walk-time
distribution $\f\left(\t\right)$ features a heavy tail that scales
as $\pro\t^{-1-\b}$ for large $\t$, with $1<\b<2$ called the ``order''
of the L\'evy walk. While the mean walk-time $\left\langle \t\right\rangle $
remains finite, naively suggesting a diffusive motion consisting of
short walks of duration $\pro\left\langle \t\right\rangle $, the
divergence of $\f\left(\t\right)$'s second moment $\left\langle \t^{2}\right\rangle $
signals the existence of unbounded fluctuations that occasionally
lead the walker on very long walks, ultimately yielding superdiffusive
dynamics. For comparison, when $\b>2$ the second moment $\left\langle \t^{2}\right\rangle $
remains finite and simple diffusion is recovered \citep{zaburdaev2015levy}.
The finite speed $v>0$ couples the walker's position to time, prevents
it from traveling a distance greater than $vt$ over a time $t$ and
guarantees the locality of its dynamics. This seemingly-innocent property
significantly complicates the L\'evy walk's analysis compared to
other superdiffusive models, like the L\'evy flight and the continuous
time random walk (CTRW), which are often easier to analyze but have
also been criticized for their non-local dynamics \citep{mantegna1994stochastic,zaburdaev2015levy}.

The approach of superdiffusive systems towards their long-time asymptotic
form was recently investigated using the L\'evy walk of order $1<\b<2$
\citep{Miron2020}. The walker's probability density $P\left(x,t\right)$
was studied for large $t$ yet \textit{beyond} the known asymptotic
solution $P_{0}\left(x,t\right)=t^{-1/\b}f\left(t^{-1/\b}x\right)$
\citep{zaburdaev2015levy,Miron2020}. This approach, captured by the
leading correction to $P_{0}\left(x,t\right)$ as $t\ra\infty$, was
shown to transition at $\b_{c}=3/2$ between a diffusive scaling $\left|x\right|\pro t^{1/2}$
for $\b>\b_{c}$ and a superdiffusive scaling $\left|x\right|\pro t^{1/\left(2\b-1\right)}$
for $\b<\b_{c}$. This transition \footnote{Not to be confused with the term ``phase transition'', that is typically
used in the context of critical phenomena.} is regarded universal as it was shown to be insensitive to $\f\left(\t\right)$'s
short-time behavior, depending only on its heavy tail $\pro\t^{-1-\b}$
\citep{Miron2020}. Indeed, recent results concerning anomalous transport
in a class of 1D systems \citep{miron2019derivation} modeled by a
L\'evy walk of order $\b=5/3$ \citep{cipriani2005anomalous,spohn2014nonlinear,PhysRevE.100.012106}
are consistent with the diffusive correction predicted in L\'evy
walks for $\b>\b_{c}$ \citep{Miron2020}. This raises the exciting
possibility that L\'evy walks may remarkably remain a valid description
of superdiffusive phenomena, even \textit{beyond} the asymptotic limit.
Elucidating how superdiffusive systems approach their asymptotic behavior
thus carries both a theoretical appeal as well as concrete consequences
for experimental and numerical investigations of superdiffusive phenomena,
which are inherently limited to finite space and time \citep{Miron2020}.
Still, one pressing question remains unanswered: what is the origin
of this transition? 

In this Rapid Communication, we investigate the mechanism responsible
for the universal transition observed in the onset of superdiffusion
in L\'evy walks \citep{Miron2020}. We find it to be twofold, consisting
of the finite speed $v$ which couples the walker's position to time
and a corresponding transition at $\b_{c}=3/2$ in the fluctuations
$\left\langle \D n_{t}^{2}\right\rangle \equiv\left\langle n_{t}^{2}\right\rangle -\left\langle n_{t}\right\rangle ^{2}$
of the number of walks $n$ completed by the walker at time $t$.
At large $t$ these become $\left\langle \D n_{t}^{2}\right\rangle \asy\left\langle \D n_{t}^{2}\right\rangle _{0}+\left\langle \d n_{t}^{2}\right\rangle $
with the asymptotic fluctuations $\left\langle \D n_{t}^{2}\right\rangle _{0}=\k_{0}t^{3-\b}$
interpolating between a ballistic scaling for $\b=1$ and a diffusive
scaling for $\b=2$ . Yet going \textit{beyond} the asymptotic limit
reveals a transition in the pre-asymptotic fluctuations $\left\langle \d n_{t}^{2}\right\rangle \asy\k_{1}t+\k_{2}t^{4-2\b}$.
For $\b>\b_{c}$ one finds $\left\langle \d n_{t}^{2}\right\rangle \propto t$,
as expected in simple diffusion (i.e. for $\b>2$). For $\b<\b_{c}$,
however, one instead finds a superdiffusive scaling $\left\langle \d n_{t}^{2}\right\rangle \pro t^{4-2\b}$.
The transition in $\left\langle \d n_{t}^{2}\right\rangle $ enters
the L\'evy walker's position through its coupling to time via $v$,
inducing a corresponding transition in the onset of superdiffusion
in the L\'evy walk propagator, hereby causally tying the two transitions.
Yet, clearly, these fluctuations may only affect models with local
dynamics, where the distance traveled by the particle is proportional
to the traveling time. As such, we complete the picture by explicitly
demonstrating the \textit{absence} of a transition in the onset of
superdiffusion in the L\'evy flight and CTRW models, where the lack
of a coupling between the particle's position and time yields non-local
dynamics. Besides explaining the onset of superdiffusion, the transition
in $\left\langle \d n_{t}^{2}\right\rangle $ also provides a tractable
observable that can be used to probe the value of $\b$ by tracking
the number of typical ``ballistic'' excursions in superdiffusive
experimental and simulation data.

Evidently, the observable $\left\langle \D n_{t}^{2}\right\rangle $
also carries significant interest in the context of ``renewal processes''
\citep{Godreche2001,Eli}, which describe physical scenarios where
the time-intervals between events are modeled as independent and identically
distributed random variables. When these intervals happen to be drawn
from a heavy-tailed distribution, with the same tail behavior $\pro\t^{-1-\b}$
as considered above, the fluctuations in the number of events are
analogous to the number of walks performed by the L\'evy walker,
similarly spreading as $\left\langle \D n_{t}^{2}\right\rangle _{0}\pro t^{3-\b}$
for asymptotically long times. Such behavior has been linked to blinking
quantum dots \citep{doi:10.1063/1.2102903}, as well as in the diffusion
of particles in polymer networks \citep{Edery} and on cell membranes
\citep{Weron2017}. The interest in the transition in $\left\langle \d n_{t}^{2}\right\rangle $,
which should similarly appear in such processes for $1<\b<2$, is
thus expected to extend far beyond the context of the onset of superdiffusion
in L\'evy walks.

\textit{The Model} - The 1D L\'evy walk of order $\b$ describes
the evolution of a walker along the infinite line in a series of uncorrelated
walks \citep{Klafter1987,zaburdaev2015levy}. In each walk, the walker
randomly draws a direction $\pm1$ and a walk-time $\t$ from the
walk-time distribution $\f\left(\t\right)$, and proceeds to walk
along the chosen direction with velocity $\pm v$ until $\t$ expires
and the process repeats. These dynamics become superdiffusive when
$\f\left(\t\right)$ features a heavy tail that scales as $\pro\t^{-1-\b}$
for large $\t$ and $1<\b<2$. In what follows we shall consider the
convenient choice
\begin{equation}
\f\left(\t\right)=\b t_{0}^{\b}\th\left[\t-t_{0}\right]\t^{-\left(1+\b\right)}\text{ }\text{for}\text{ }1<\b<2,\label{eq:phi}
\end{equation}
where the step function $\th\left[x\right]$ keeps $\f\left(\t\right)$
normalizable on $\t\in\co{0,\infty}$ by imposing a cutoff at the
minimal walk time $t_{0}$. For simplicity, however, we shall henceforth
set $t_{0}=1$, effectively rendering $\t$ to be a dimensionless
time.

\begin{figure}
\begin{centering}
\includegraphics[scale=0.46]{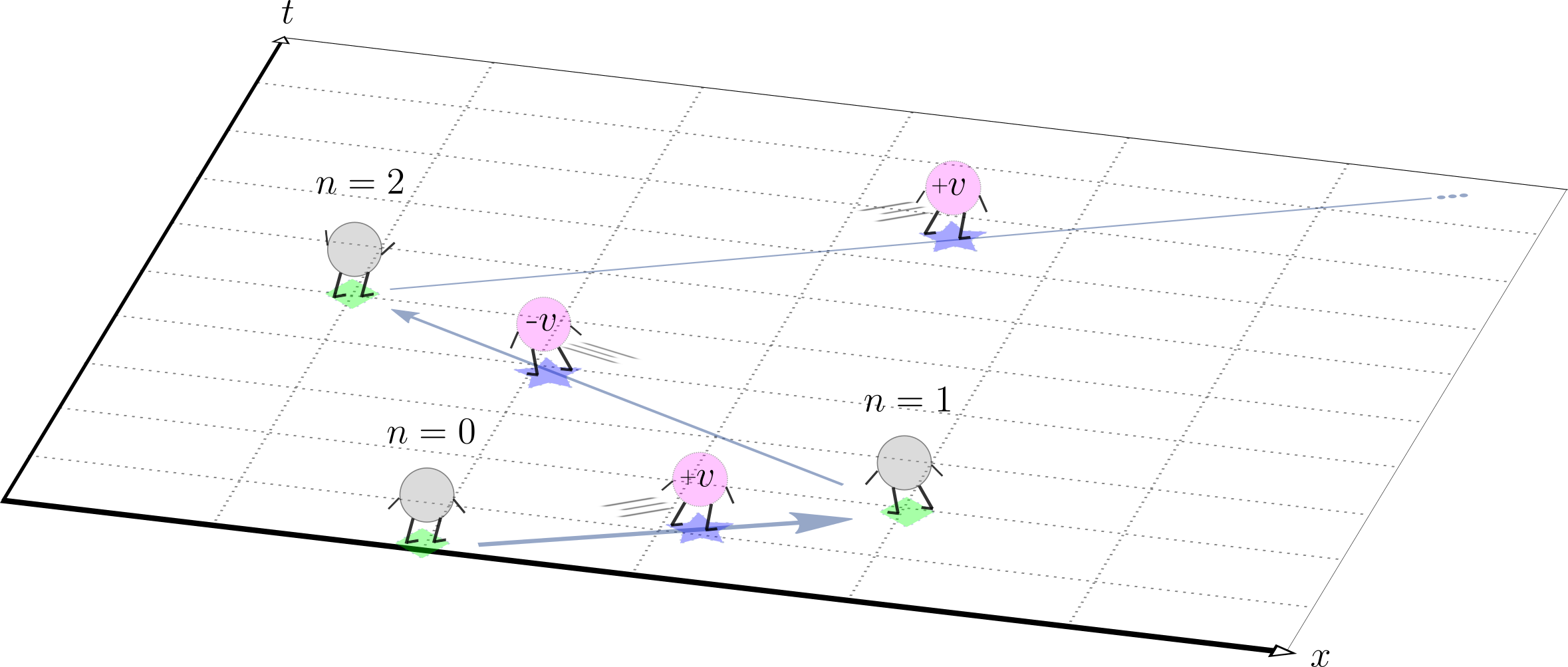}
\par\end{centering}
\caption{A schematic illustration of the first few steps of the L\'evy walk
dynamics. Green diamonds mark the walker's position after \textit{completing
}its $n$'th walk, as described by $\protect\n_{n}\left(x,t\right)$
in Eq. (\ref{eq:nu_eqn}). Purple stars mark the walker's position
\textit{during} its $n$'th walk, as described by $P_{n}\left(x,t\right)$
in Eq. (\ref{eq:P_eqn}).}
\label{fig: illustration}
\end{figure}

To study the fluctuations in the number of steps $n$ completed by
the walker at time $t$, we generalize the L\'evy walk model \citep{zaburdaev2015levy}
and formulate self-consistent equations for two quantities: the density
per unit-time $\n_{n}\left(x,t\right)$ of walkers \textit{leaving}
position $x$ at time $t$ after completing $n$ walks and the density
$P_{n}\left(x,t\right)$ of walkers \textit{at} position $x$ at time
$t$ during their $n$'th walk. The equations for $\n_{n}\left(x,t\right)$
and its initial condition are
\[
\n_{n+1}\left(x,t\right)=\frac{1}{2}\int_{-\infty}^{\infty}\dif y\int_{0}^{t}\dif\t\d\left[\left|y\right|-v\t\right]
\]
\begin{equation}
\times\f\left(\t\right)\n_{n}\left(x-y,t-\t\right)\text{ and }\n_{0}\left(x,t\right)=\d\left(x\right)\d\left(t\right).\label{eq:nu_eqn}
\end{equation}
The right-hand side accounts for contributions to $\n_{n+1}\left(x,t\right)$
from walkers that have just completed $n$ steps and are located at
$x-y$ at time $t-\t$. The equation for $P_{n}\left(x,t\right)$
is given by \footnote{Note that Eq. (\ref{eq:P_eqn}) for $P_{n}\left(x,t\right)$ does
not require an initial condition beyond that provided in Eq. (\ref{eq:nu_eqn})
for $\n_{n}\left(x,t\right)$.}
\[
P_{n}\left(x,t\right)=\frac{1}{2}\int_{-\infty}^{\infty}\dif y\int_{0}^{t}\dif\t
\]
\begin{equation}
\times\d\left[\left|y\right|-v\t\right]\p\left(\t\right)\n_{n}\left(x-y,t-\t\right),\label{eq:P_eqn}
\end{equation}
describing contributions to $P_{n}\left(x,t\right)$ from walkers
beginning their $n$'th walk at position $x-y$ at time $t-\t$ and
drawing a walk-time \textit{greater} than $\t$, which occurs with
probability 
\begin{equation}
\p\left(\t\right)=\int_{\t}^{\infty}\dif u\f\left(u\right).\label{eq:psi}
\end{equation}
These walkers pass through $x$ at time $t$, \textit{before} the
walk's duration had expired, and then continue walking. Since we consider
a $\d\left(x\right)$ initial condition and the number of walkers
is conserved, $P_{n}\left(x,t\right)$ is hereafter referred to as
the `` generalized propagator'' and may be thought of as a probability
density of a single walker. The model's dynamics and the differences
between $\n_{n}\left(x,t\right)$ and $P_{n}\left(x,t\right)$ are
illustrated in Fig. \ref{fig: illustration}.

\begin{figure}
\includegraphics[scale=0.55]{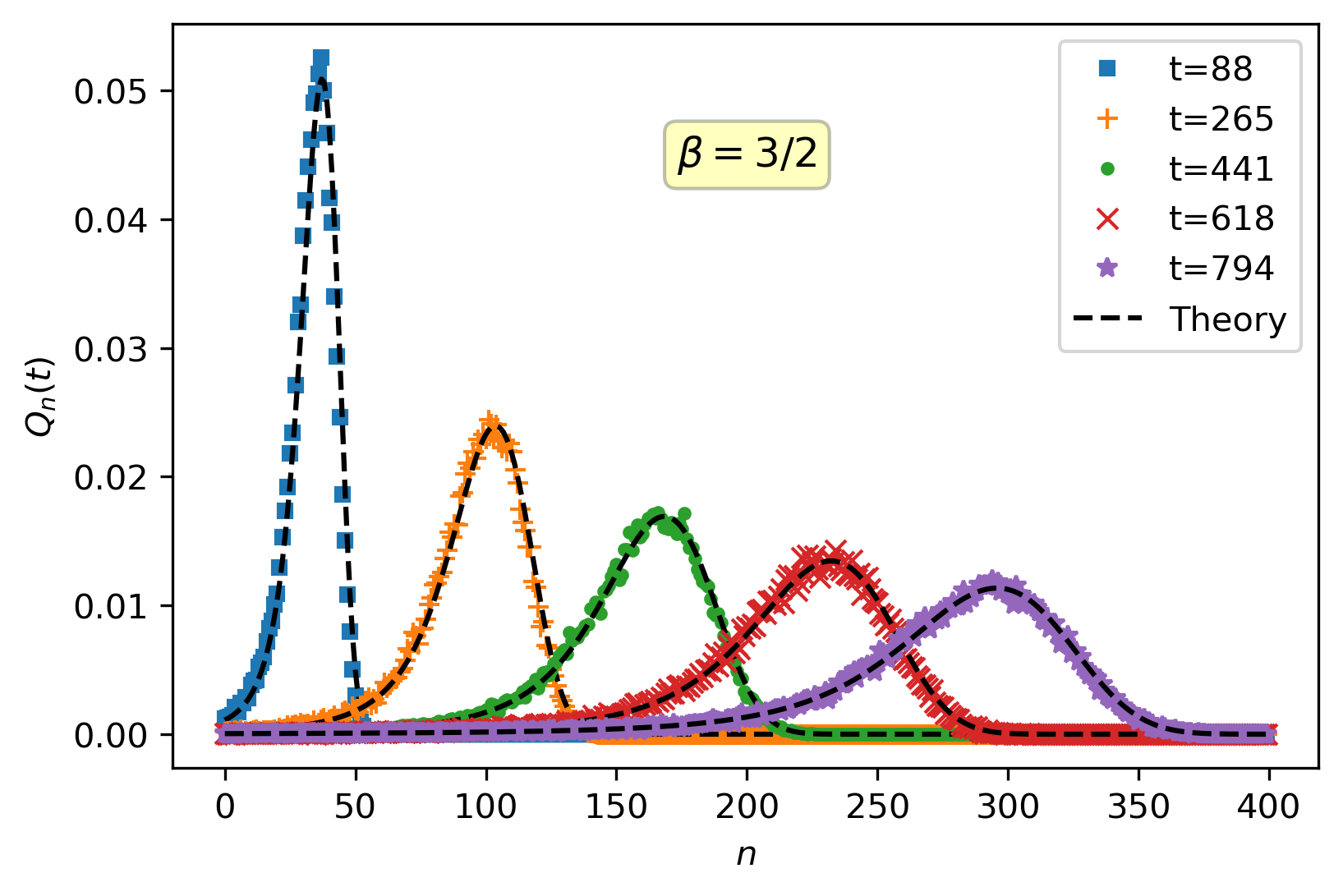}

\caption{The walk-number distribution $Q_{n}\left(t\right)$ for $\protect\b=\protect\b_{c}$
versus the number of steps $n$. Markers depict the simulated walk-number
distribution while the dashed black lines represent the numerical
inverse Laplace-transform of $\tilde{Q}_{n}\left(s\right)$ in Eq.
(\ref{eq:Q_n_s}) at different times. The temporal growth of the distribution's
width is described by $\left\langle \protect\D n_{t}^{2}\right\rangle $.}
\label{fig 2}
\end{figure}

\begin{figure*}
\begin{centering}
\includegraphics[scale=0.55]{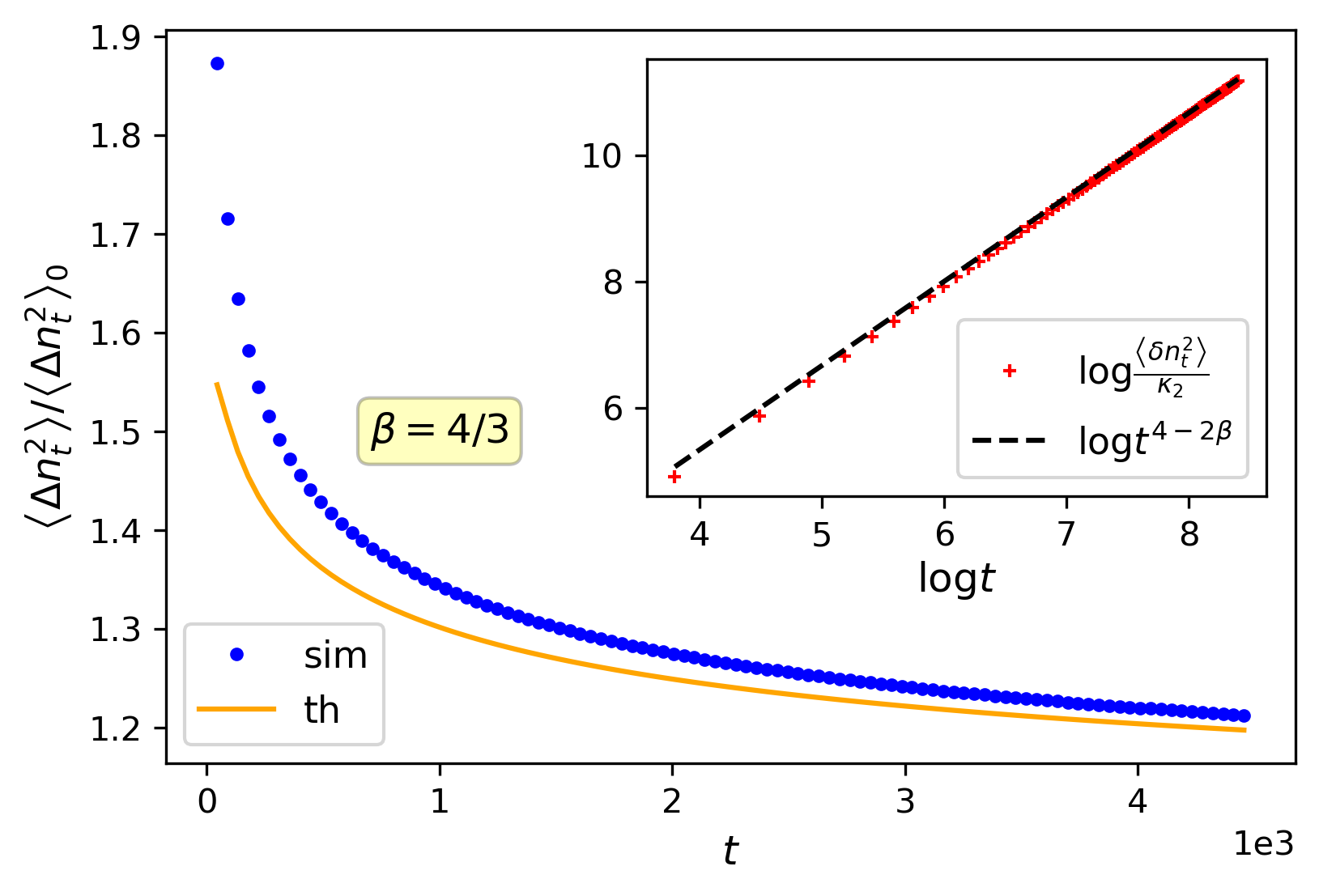}\hfill{}\includegraphics[scale=0.55]{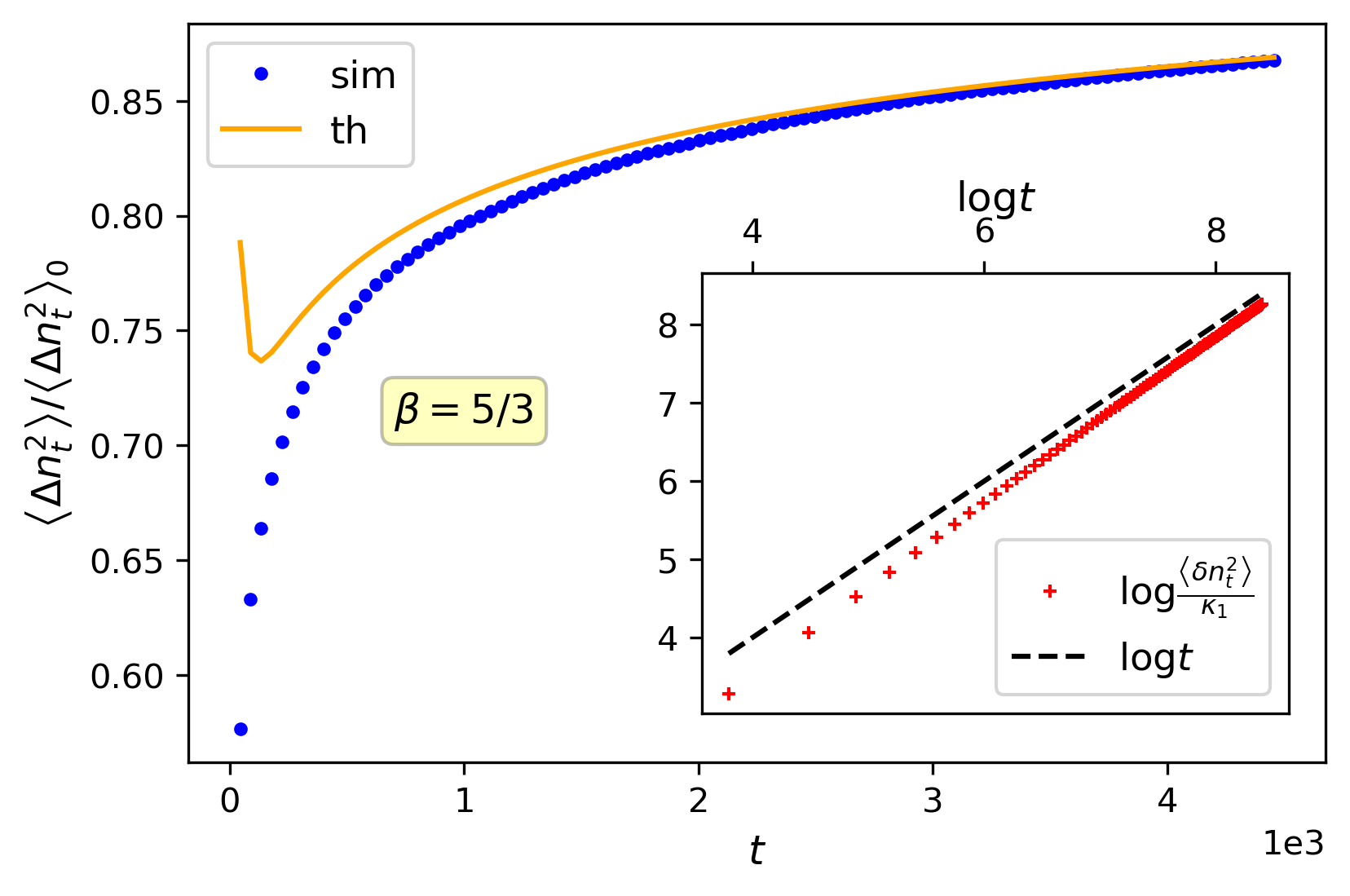}
\par\end{centering}
\caption{The large-$t$ fluctuations $\left\langle \protect\D n_{t}^{2}\right\rangle $
over the asymptotic fluctuations $\left\langle \protect\D n_{t}^{2}\right\rangle _{0}$
of Eq. (\ref{eq:Delta n_t^2}) versus $t$ . As $t\protect\ra\infty$,
this ratio approaches unity from above for $\protect\b<\protect\b_{c}$
(left) and from below for $\protect\b>\protect\b_{c}$ (right). Blue
circles depict the simulated $\left\langle \protect\D n_{t}^{2}\right\rangle $
while the solid orange curve depicts its theoretical expression of
Eq. (\ref{eq:Delta n_t^2}). The insets show a logarithmic plot of
the pre-asymptotic fluctuations $\left\langle \protect\d n_{t}^{2}\right\rangle =\protect\k_{1}t+\protect\k_{2}t^{4-2\protect\b}$
versus $t$, where $\left\langle \protect\d n_{t}^{2}\right\rangle $
is obtained by subtracting the theoretical $\left\langle \protect\D n_{t}^{2}\right\rangle _{0}$
from the simulated fluctuations $\left\langle \protect\D n_{t}^{2}\right\rangle $.
Specifically, the left panel depicts $\log\left\langle \protect\d n_{t}^{2}\right\rangle /\protect\k_{2}$
by red crosses and $\left(4-2\protect\b\right)\log t$ by a black
dashed curve for $\protect\b=4/3<\protect\b_{c}$ while the right
panel uses the same scheme to depict $\log\left\langle \protect\d n_{t}^{2}\right\rangle /\protect\k_{1}$
and $\log t$ for $\protect\b=5/3>\protect\b_{c}$.}

\label{fig 3}
\end{figure*}

\textit{Main Results} - The universal transition in the onset of superdiffusion
in L\'evy walks \citep{Miron2020} is shown to trace back to a corresponding
transition in the fluctuations $\left\langle \D n_{t}^{2}\right\rangle $
of the number of walks $n$ completed by the walker at time $t$.
These fluctuations, and the entailing transition, then enter the walker's
position through its coupling to time via $v$, inducing a corresponding
transition in the L\'evy walk propagator. We first compute the generalized
propagator $P_{n}\left(x,t\right)$, from which we derive an exact
expression for the walk-number distribution $\tilde{Q}_{n}\left(s\right)$
in Laplace-space in Eq. (\ref{eq:Q_n_s}). The real-time distribution
$Q_{n}\left(t\right)$ is verified against direct numerical simulations
in Fig. \ref{fig 2}. We next explicitly evaluate the large-$t$ walk-number
fluctuations in Eq. (\ref{eq:Delta n_t^2}), finding
\begin{equation}
\left\langle \D n_{t}^{2}\right\rangle \asy\left\langle \D n{}_{t}^{2}\right\rangle _{0}+\left\langle \d n_{t}^{2}\right\rangle ,\label{eq:n_t fluct}
\end{equation}
where $\left\langle \D n{}_{t}^{2}\right\rangle _{0}=\k_{0}t^{3-\b}$
describes the asymptotic behavior while $\left\langle \d n_{t}^{2}\right\rangle =\k_{1}t+\k_{2}t^{4-2\b}$
accounts for the pre-asymptotic fluctuations that, as shown in Fig.
\ref{fig 3}, undergo a transition at $\b_{c}=3/2$ with $\k_{0}$,
$\k_{1}$ and $\k_{2}$ given in Eq. (\ref{eq:kappa}). Specifically,
for $\b>\b_{c}$ we recover diffusive fluctuations $\left\langle \d n_{t}^{2}\right\rangle \propto t$
whereas, for $\b<\b_{c}$, the pre-asymptotic fluctuations instead
grow superdiffusively as $\left\langle \d n_{t}^{2}\right\rangle \pro t^{4-2\b}$.
Clearly, fluctuations in $n$ may affect the propagator \textit{only
if} the distance traveled by the particle is proportional to the traveling
duration. Nevertheless, we complete the picture by demonstrating the
absence of a transition in the onset of superdiffusion in the L\'evy
flight and CTRW models, where the particle's position is \textit{not}
coupled to time. Consequently, although a transition in $\left\langle \d n_{t}^{2}\right\rangle $
is found in the CTRW model, it fails to induce a corresponding transition
in the onset of superdiffusion.

\textit{The Generalized Propagator} - Applying a Fourier-Laplace transform
to Eq. (\ref{eq:nu_eqn}) for $\n_{n}\left(x,t\right)$ yields
\[
\til{\n}_{n+1}\left(k,s\right)=\frac{1}{2}\int_{0}^{\infty}\dif te^{-st}
\]
\begin{equation}
\times\int_{0}^{t}\dif\t\f\left(\t\right)\left(e^{ivk\t}+e^{-ivk\t}\right)\hat{\n}_{n}\left(k,t-\t\right),\label{eq:nu_eqn_1}
\end{equation}
where we denote the Fourier transform by $\hat{f}\left(k,t\right)=\int_{-\infty}^{\infty}\dif xe^{-ikx}f\left(x,t\right)$
and the Laplace transform by $\til f\left(k,s\right)=\int_{0}^{\infty}\dif te^{-st}\hat{f}\left(k,t\right)$.
Interchanging the order of integration of $t$ and $\t$, i.e. $\int_{0}^{\infty}\dif t\int_{0}^{t}\dif\t\ra\int_{0}^{\infty}\dif\t\int_{\t}^{\infty}\dif t$,
lets us reduce Eq. (\ref{eq:nu_eqn_1}) to
\begin{equation}
\til{\n}_{n+1}\left(k,s\right)=\frac{1}{2}\left[\til{\f}\left(s-ivk\right)+\til{\f}\left(s+ivk\right)\right]\tilde{\n}_{n}\left(k,s\right).\label{eq:nu_eqn_2}
\end{equation}
Solving the $n$ dependence in Eq. (\ref{eq:nu_eqn_2}) subject to
the initial condition in Eq. (\ref{eq:nu_eqn}) gives
\begin{equation}
\tilde{\n}_{n}\left(k,s\right)=\left(\frac{1}{2}\left[\til{\f}\left(s-ivk\right)+\til{\f}\left(s+ivk\right)\right]\right)^{n}.\label{eq:nu_sol}
\end{equation}
Applying the same approach to Eq. (\ref{eq:P_eqn}) for $P_{n}\left(x,t\right)$
leads to 
\begin{equation}
\til P_{n}\left(k,s\right)=\frac{1}{2}\til{\n}_{n}\left(k,s\right)\left[\til{\p}\left(s-ivk\right)+\til{\p}\left(s+ivk\right)\right].\label{eq:P_almost_sol}
\end{equation}
The formal solution for $\til P_{n}\left(k,s\right)$ is obtained
by combining Eqs. (\ref{eq:nu_sol}) and (\ref{eq:P_almost_sol})
into
\begin{equation}
\til P_{n}\left(k,s\right)=\frac{\til{\p}\left(s-ivk\right)+\til{\p}\left(s+ivk\right)}{2^{n+1}\left[\til{\f}\left(s-ivk\right)+\til{\f}\left(s+ivk\right)\right]^{-n}}.\label{eq:P_sol}
\end{equation}

The generalized L\'evy walk propagator $\til P_{n}\left(k,s\right)$
in Eq. (\ref{eq:P_sol}) is consistent with the known L\'evy walk
propagator $\til P^{LW}\left(k,s\right)=\frac{\til{\p}\left(s-ivk\right)+\til{\p}\left(s+ivk\right)}{2-\til{\f}\left(s-ivk\right)-\til{\f}\left(s+ivk\right)}$
\citep{zaburdaev2015levy}, which is immediately recovered when summing
$\til P_{n}\left(k,s\right)$ over $n$. For large $t$ and small
$\left|k\right|$, $\til P^{LW}\left(k,s\right)$ was shown in \citep{Miron2020}
to asymptotically approach $\hat{P}^{LW}\left(q,t\right)\asy e^{-tI\left(q\right)}$,
where $I\left(q\right)=\left(1-\text{Re}\left[\hat{\f}\left(q\right)\right]\right)/\partial_{q}\im\left[\hat{\f}\left(q\right)\right]$
and $q=vk$. There, this non-trivial functional dependence of $I\left(q\right)$
on $\hat{\f}\left(q\right)$ stems from the spatio-temporal coupling
by $v$, as dictated by the distribution $\til{\f}\left(s\pm ivk\right)$,
and ultimately yields the transition in the onset of superdiffusion
in L\'evy walks. While uncovering the origin of this transition,
we shall see that the coupling between the walk's position and time
is, in fact, a simple, natural and intuitive mechanism which serves
to intertwine the walk's position with the corresponding transition
in the walk-number fluctuations.

\textit{Walk-Number Fluctuations} - Our next task is to evaluate the
walk-number fluctuations $\left\langle \D n_{t}^{2}\right\rangle $.
To this end, we first marginalize the generalized propagator $\til P_{n}\left(k,s\right)$
over the walker's position by setting $k=0$. This, along with the
Laplace transform $\til{\p}\left(s\right)=s^{-1}\left(1-\til{\f}\left(s\right)\right)$
of $\p\left(\t\right)$ in Eq. (\ref{eq:psi}), yields the Laplace-transformed
walk-number distribution
\begin{equation}
\til Q_{n}\left(s\right)\equiv\til P_{n}\left(k=0,s\right)=s^{-1}\left(1-\til{\f}\left(s\right)\right)\til{\f}\left(s\right)^{n}.\label{eq:Q_n_s}
\end{equation}
While we here derive it from the generalized propagator $\til P_{n}\left(k,s\right)$,
we stress that $\til Q_{n}\left(s\right)$ is a more fundamental quantity
that can be obtained without considering the L\'evy walker's spatial
behavior (see \citep{SM}). To proceed, we introduce the Laplace-space
moment generating function
\begin{equation}
g\left(s;\l\right)=\sum_{n=0}^{\infty}\l^{n}\til Q_{n}\left(s\right)=\frac{1-\til{\f}\left(s\right)}{s\left(1-\l\til{\f}\left(s\right)\right)},\label{eq:mom_gen_func}
\end{equation}
from which we derive
\begin{equation}
\left\langle \tilde{n}_{s}\right\rangle =\frac{\til{\f}\left(s\right)}{s\left(1-\til{\f}\left(s\right)\right)}\text{ and }\left\langle \tilde{n}_{s}^{2}\right\rangle =\frac{\til{\f}\left(s\right)\left(1+\til{\f}\left(s\right)\right)}{s\left(1-\til{\f}\left(s\right)\right)^{2}},\label{eq:n_s moments}
\end{equation}
noting that $\left\langle \tilde{n}_{s}^{m}\right\rangle =\int_{0}^{\infty}\dif te^{-st}\left\langle n_{t}^{m}\right\rangle $
is the Laplace transform of the $m$'th moment $\left\langle n_{t}^{m}\right\rangle $.

To keep our discussion as general as possible, let us consider a generic
walk-time distribution $\f\left(\t\right)$ with an analytic short-time
behavior and a tail which scales as $\pro\t^{-1-\b}$ for large $\t$.
The Laplace transform $\til{\f}\left(s\right)=\int_{0}^{\infty}\dif te^{-st}\f\left(t\right)$
of such a general distribution is given by
\begin{equation}
\til{\f}\left(s\right)=s^{\b}\sum_{r=0}^{\infty}d_{r}s^{r}+\sum_{m=0}^{\infty}c_{m}s^{m}.\label{eq:general phi}
\end{equation}
The singular terms $s^{\b}\sum_{r=0}^{\infty}d_{r}s^{r}$ account
for the distribution's tail and are responsible for the divergence
of the second and higher moments, while the analytic series $\sum_{m=0}^{\infty}c_{m}s^{m}$
captures its short-time behavior. The coefficients $\left\{ c_{m}\right\} _{m=0}^{\infty}$
and $\left\{ d_{r}\right\} _{r=0}^{\infty}$ may be uniquely determined
for any such walk-time distribution $\f\left(\t\right)$, including
the choice in Eq. (\ref{eq:phi}) which was used in Figs. \ref{fig 2}
and \ref{fig 3} but also for other choices (see \citep{SM}). We
proceed to analyze the large-$t$ behavior of $\left\langle \D n{}_{t}^{2}\right\rangle $
by first obtaining the small-$s$ (i.e. large-$t$) behavior of $\left\langle \tilde{n}_{s}\right\rangle $
and $\left\langle \tilde{n}_{s}^{2}\right\rangle $ in Eq. (\ref{eq:n_s moments})
and then taking the inverse Laplace transform (see \citep{SM}). We
find
\begin{equation}
\left\langle n_{t}\right\rangle \approx-\frac{t}{c_{1}}+\frac{d_{0}t^{2-\b}}{\G\left[3-\b\right]c_{1}^{2}}+\frac{c_{2}-c_{1}^{2}}{c_{1}^{2}},\label{eq:av_n_t}
\end{equation}
and
\begin{equation}
\left\langle n_{t}^{2}\right\rangle \approx\frac{t^{2}}{c_{1}^{2}}-\frac{4d_{0}t^{3-\b}}{\G\left[4-\b\right]c_{1}^{3}}+\frac{\left(3c_{1}^{2}-4c_{2}\right)t}{c_{1}^{3}}+\frac{6d_{0}t^{4-2\b}}{\G\left[5-2\b\right]c_{1}^{4}},\label{eq:av_n_t_sqr}
\end{equation}
where higher order terms are neglected and the normalization condition
$c_{0}=1$ is used. The leading large-$t$ behavior of $\left\langle \D n_{t}^{2}\right\rangle \approx\left\langle \D n{}_{t}^{2}\right\rangle _{0}+\left\langle \d n_{t}^{2}\right\rangle $
in Eq. (\ref{eq:n_t fluct}) is thus 
\begin{equation}
\left\langle \D n{}_{t}^{2}\right\rangle _{0}=\k_{0}t^{3-\b}\text{ and }\left\langle \d n_{t}^{2}\right\rangle =\k_{1}t+\k_{2}t^{4-2\b},\label{eq:Delta n_t^2}
\end{equation}
with the coefficients $\k_{0},\k_{1}$ and $\k_{2}$ given by
\[
\k_{0}=\frac{2\left(1-\b\right)d_{0}}{\G\left[4-\b\right]c_{1}^{3}},\text{ }\k_{1}=\frac{c_{1}^{2}-2c_{2}}{c_{1}^{3}}
\]
\begin{equation}
\text{ and }\k_{2}=\frac{d_{0}}{c_{1}^{4}}\left(\frac{6}{\G\left[5-2\b\right]}-\frac{1}{\G\left[3-\b\right]^{2}}\right).\label{eq:kappa}
\end{equation}
For the choice of $\f\left(\t\right)$ in Eq. (\ref{eq:phi}), one
finds $c_{1}=-\frac{\b}{\b-1}$, $d_{0}=-\G\left[1-\b\right]$ and
$c_{2}=-\frac{\b}{2\left(2-\b\right)}$ (see \citep{SM}). These walk-number
fluctuations enter the L\'evy walk propagator since the distance
traveled by the walker is proportional to its traveling time. As such,
the transition in the pre-asymptotic fluctuations $\left\langle \d n_{t}^{2}\right\rangle $
induces a corresponding transition in the onset of superdiffusion.

\textit{The }L\'evy\textit{ flight and CTRW models} - We finally
demonstrate the \textit{absence} of a transition in the onset of superdiffusion
in the L\'evy flight and CTRW models, where the distance traveled
by the particle is not proportional to the traveling time. Moreover,
we explicitly show that the CTRW's non-local dynamics fail to produce
a transition in the onset of superdiffusion, even though the model
does exhibit a transition in $\left\langle \d n_{t}^{2}\right\rangle $.
Explicit calculations and details are provided in \citep{SM}.

In each step of the 1D CTRW dynamics, the particle waits a random
time $\t$ and then makes a random jump $\el$ \citep{zaburdaev2015levy}.
Superdiffusion arises when the waiting-time distribution scales as
$\o\left(\t\right)\pro\t^{-1-\b}$ for large $\t$ and has a finite
first moment $\left\langle \t\right\rangle $, corresponding to $\b>1$,
while the symmetric jump-distance distribution scales as $g\left(\el\right)\pro\left|\el\right|^{-1-\g}$
for large $\left|\el\right|$ and has a diverging second moment $\left\langle \el^{2}\right\rangle \ra\infty$,
corresponding to $1<\g<2$. Generalizing the CTRW dynamics to account
for the number of steps $n$, as in Eqs. (\ref{eq:nu_eqn}) and (\ref{eq:P_eqn})
for the L\'evy walk model, one obtains the generalized CTRW propagator
\begin{equation}
\tilde{P}_{n}^{CTRW}\left(k,s\right)=s^{-1}\left(1-\tilde{\o}\left(s\right)\right)\hat{g}\left(k\right)^{n}\tilde{\o}\left(s\right)^{n}.\label{eq:CTRW propagator}
\end{equation}
Marginalizing over space gives the walk-number distribution $\tilde{Q}_{n}^{CTRW}\left(s\right)=s^{-1}\left(1-\tilde{\o}\left(s\right)\right)\tilde{\o}\left(s\right)^{n}$,
which is identical to that obtained in Eq. (\ref{eq:Q_n_s}) for the
L\'evy walk. As such, the same transition arises in the pre-asymptotic
walk-number fluctuations $\left\langle \d n_{t}^{2}\right\rangle $
at $\b_{c}$, as in Eq. (\ref{eq:Delta n_t^2}). However, by marginalizing
Eq. (\ref{eq:CTRW propagator}) over $n$ and taking the long time
and large distance limit, one finds $\tilde{P}^{CTRW}\left(k,t\right)\asy e^{-t\left(\bar{D}_{0}\left|k\right|^{\g}-\overline{D}_{1}k^{2}+\ord{\left|k\right|^{2+\g}}\right)}$,
where $\bar{D}_{0}$ and $\overline{D}_{1}$ depend on the details
of $\o\left(\t\right)$ and $g\left(\el\right)$. Since the leading
correction to the asymptotic CTRW propagator $\tilde{P}_{0}^{CTRW}\left(k,t\right)=e^{-\bar{D}_{0}t\left|k\right|^{\g}}$
is proportional to $\sim k^{2}$ for \textit{any} $1<\g<2$, no transition
arises in the onset of superdiffusion. 

A similar picture is found in the 1D L\'evy flight, which describes
a ``flier'' whose discrete evolution consists of repeatedly drawing
a flight-distance $\el$ from the distribution $\x\left(\el\right)$
and immediately materializing at its new location. Superdiffusion
appears when $\x\left(\el\right)$'s symmetric tails scale as $\x\left(\el\right)\pro\left|\el\right|^{-1-\b}$
for large $\left|\el\right|$ and $1<\b<2$. The model's discrete
evolution is neatly contained within the generalized L\'evy walk
dynamics of Eqs. (\ref{eq:nu_eqn}) and (\ref{eq:P_eqn}) and its
known propagator $\hat{P}_{n}^{LF}\left(k\right)=\til{\f}\left(k\right)^{n}$
is recovered from $\hat{\n}_{n}\left(k,s\right)$ of Eq. (\ref{eq:nu_sol})
by setting $s=0$. For large distances and $n$, it assumes the form
$\hat{P}_{n}^{LF}\left(k\right)\asy e^{-n\left(D_{0}\left|k\right|^{\b}-D_{1}k^{2}+\ord{k^{4}}\right)}$,
where $D_{0}$ and $D_{1}$ depend on the details of $\x\left(\el\right)$.
Again, no transition appears in the onset of superdiffusion.

\textit{Conclusions} - In this paper, we studied the mechanism behind
the recently reported universal transition in the onset of superdiffusion
in L\'evy walks of order $1<\b<2$. It was shown to be twofold, consisting
of the finite speed $v$ which couples the walker's position to time
and a corresponding transition in the fluctuations of the number of
walks $n$ completed by the walker at time $t$. Generalizing the
L\'evy walk model to account for the number of walks $n$ allowed
us to compute the walk-number distribution and its large-$t$ fluctuations
$\left\langle \D n_{t}^{2}\right\rangle \asy\left\langle \D n_{t}^{2}\right\rangle _{0}+\left\langle \d n_{t}^{2}\right\rangle $.
A transition was demonstrated in the pre-asymptotic fluctuations $\left\langle \d n_{t}^{2}\right\rangle =\k_{1}t+\k_{2}t^{4-2\b}$,
showing diffusive behavior $\left\langle \d n_{t}^{2}\right\rangle \propto t$
for $\b>\b_{c}$ and superdiffusive behavior $\left\langle \d n_{t}^{2}\right\rangle \pro t^{4-2\b}$
for $\b<\b_{c}$. This picture was completed by showing that no transition
occurs in the onset of superdiffusion in the L\'evy flight and CTRW
models, where the particle's position is not coupled to time.

Unlike the full propagator, which is nutritiously hard to obtain from
data, the walk-number fluctuations $\left\langle \D n_{t}^{2}\right\rangle $
can readily be extracted from the dynamics by tracking the evolution
of the number of typical ``ballistic'' excursions observed in superdiffusive
systems. This study shows that this robust and accessible observable
can be used to precisely predict which systems are expected to exhibit
a transition in the onset of superdiffusion, be they experimental
or numerical. However, besides its theoretical value in uncovering
the mechanism responsible for the transition in the onset of superdiffusion
\citep{Miron2020}, the transition in $\left\langle \D n_{t}^{2}\right\rangle $
can itself be used as a tool for precisely determining the value of
$\b$. This collateral contribution is important since only a few
such instruments are currently known, in spite of the well-known and
often devastating difficulties posed by finite-time corrections in
both experimental and numerical studies of superdiffusive phenomena
\citep{cipriani2005anomalous,benhamou2007many,sims2007minimizing,gonzalez2008understanding,harris2012generalized,PhysRevLett.112.110601,PhysRevE.100.042140,agrawal2020anomalous}.
This work joins the efforts detailed in \citep{kessler2010infinite,dechant2011solution,hazut2015fractional,miron2019derivation,PhysRevE.100.012106,Miron2020}
of establishing an understanding of the pre-asymptotic behavior of
superdiffusive systems. In this context, it would be very interesting
to test these predictions in experimental and numerical systems which
are modeled by L\'evy walks.

\textit{Acknowledgments} - I thank David Mukamel for his ongoing encouragement
and support, for critically reading this manuscript and for many helpful
discussions. I also thank Julien Cividini and Oren Raz for critically
reading this manuscript and for their helpful remarks. This work was
supported by a research grant from the Center of Scientific Excellence
at the Weizmann Institute of Science.

\cleardoublepage{}

\section*{Supplemental Material - The origin of universality in the onset of
superdiffusion in L\'evy walks}

\section{Space-Independent Derivation of $\protect\til Q_{n}\left(s\right)$}

\begin{figure*}
\begin{centering}
\includegraphics[scale=0.55]{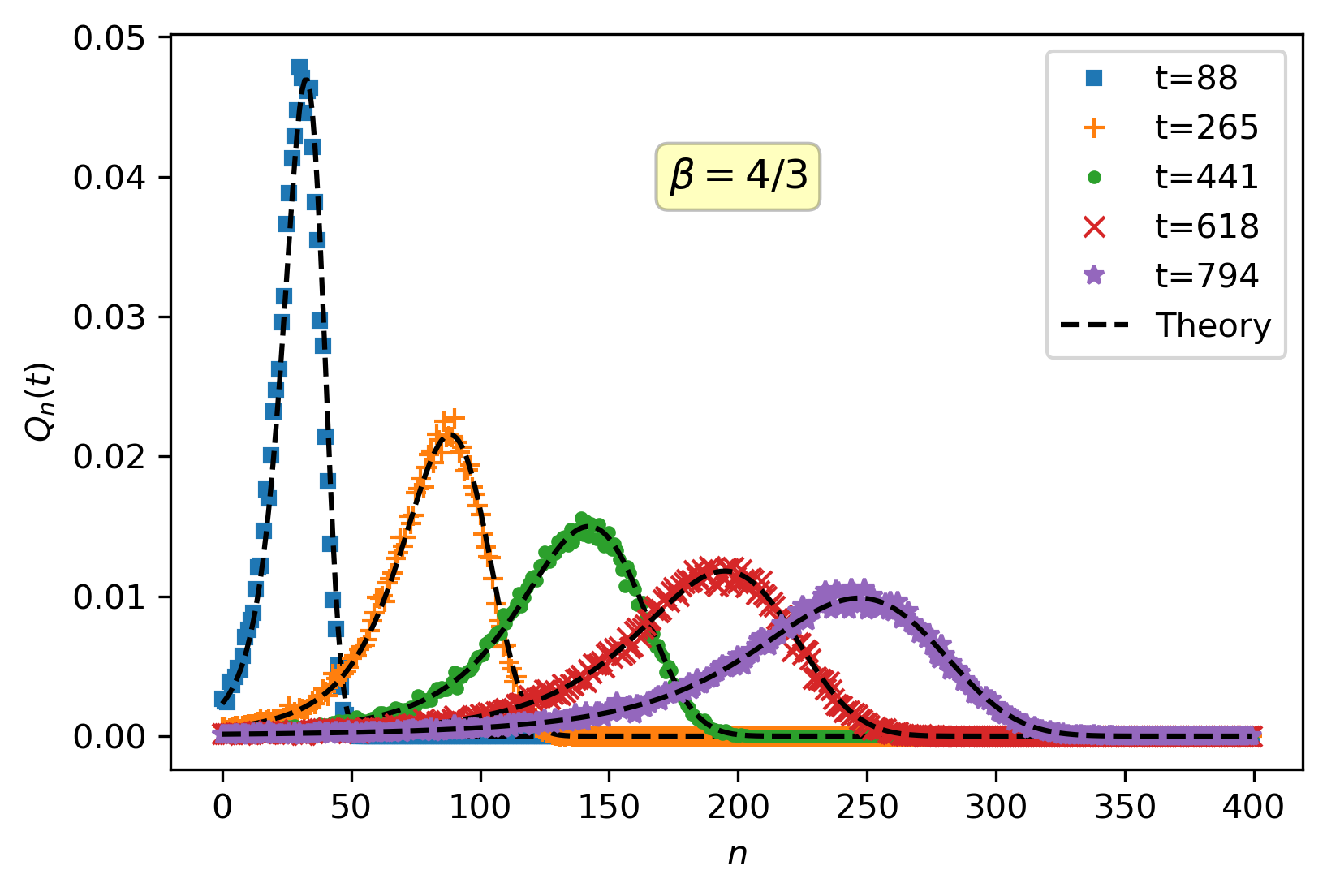} \hfill{}
\includegraphics[scale=0.55]{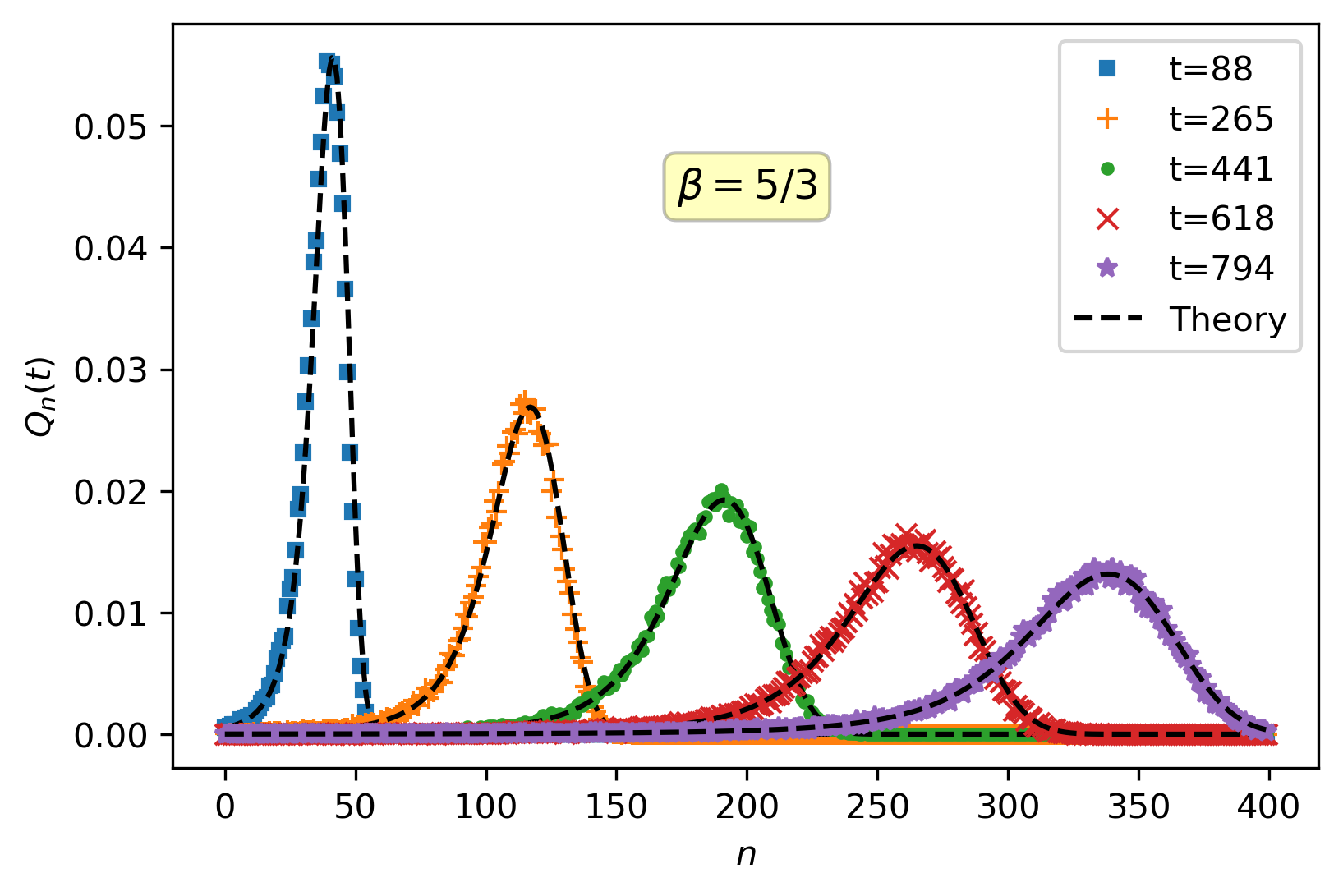}
\par\end{centering}
\caption{The walk-number distribution $Q_{n}\left(t\right)$ versus the number
of steps $n$. Markers depict the simulated walk-number distribution
while the dashed black lines represent the numerical inverse Laplace-transform
of $\tilde{Q}_{n}\left(s\right)$ in Eq. (\ref{eq:Q_n(s) sol}) at
different times. The left panel shows $\protect\b=4/3<\protect\b_{c}$
while the right panel shows $\protect\b=5/3>\protect\b_{c}$. One
may appreciate the different temporal growth of the distribution's
width, described by $\left\langle \protect\D n_{t}^{2}\right\rangle $,
for different values of $\protect\b$.}

\label{Q_n(t) fig}
\end{figure*}

In this section we detail a space-independent derivation of the walk-number
distribution. Just as we did for the generalized propagator in the
main text Eq. $\left(10\right)$, one may write down self-consistent
equations which ultimately yield the Laplace-space walk-number distribution
$\til Q_{n}\left(s\right)$ obtained in the main text Eq. $\left(11\right)$.
To this end, we define the density per unit time $R_{n}\left(t\right)$
of walkers completing their $n$'th walk at time $t$ and the density
of walkers $Q_{n}\left(t\right)$ that are in the midst of their $n$'th
walk at time $t$. The equation and initial condition for $R_{n}\left(t\right)$
are 
\begin{equation}
R_{n+1}\left(t\right)=\int_{0}^{t}\dif\t R_{n}\left(t-\t\right)\f\left(\t\right)\text{ and }R_{0}\left(t\right)=1,\label{eq:R(t)}
\end{equation}
while the equation for $Q_{n}\left(t\right)$ is
\begin{equation}
Q_{n}\left(t\right)=\int_{0}^{t}\dif\t R_{n}\left(t-\t\right)\psi\left(\t\right),\label{eq:Q(t)}
\end{equation}
where $\psi\left(t\right)=\int_{t}^{\infty}\dif\t\f\left(\t\right)$
is the probability of drawing a walk-time greater than $t$. Taking
the Laplace transform $\tilde{f}\left(s\right)=\int_{0}^{\infty}\dif te^{-st}f\left(t\right)$
of Eq. (\ref{eq:R(t)}) and interchanging the integration limits $\int_{0}^{\infty}\dif t\int_{0}^{t}\dif\t\ra\int_{0}^{\infty}\dif\t\int_{\t}^{\infty}\dif t$
yields
\begin{equation}
\tilde{R}_{n+1}\left(t\right)=\tilde{R}_{n}\left(s\right)\tilde{\f}\left(s\right).\label{eq:R(s) eq}
\end{equation}
Using the initial condition in Eq. (\ref{eq:R(t)}), $\tilde{R}_{n}\left(s\right)$
is solved by 
\begin{equation}
\tilde{R}_{n}\left(s\right)=\tilde{\f}\left(s\right)^{n}.\label{eq:R(s) sol}
\end{equation}
Repeating the same procedure for $Q_{n}\left(t\right)$ and using
$\tilde{\psi}\left(s\right)=s^{-1}\left(1-\tilde{\f}\left(s\right)\right)$
yields
\begin{equation}
\tilde{Q}_{n}\left(s\right)=s^{-1}\left(1-\tilde{\f}\left(s\right)\right)\tilde{\f}\left(s\right)^{n},\label{eq:Q_n(s) sol}
\end{equation}
which is the same distribution found in the main text Eq. $\left(11\right)$
by marginalizing the generalized L\'evy walk propagator over space.
Figure \ref{Q_n(t) fig} illustrates the vastly different spreading
of $Q_{n}\left(t\right)$, captured by $\left\langle \D n_{t}^{2}\right\rangle $,
for $\b=4/3<\b_{c}$ and $\b=5/3>\b_{c}$.

\section{$\left\langle n_{t}\right\rangle $ and $\left\langle n_{t}^{2}\right\rangle $}

In this section we detail the derivation of $\left\langle n_{t}\right\rangle $
and $\left\langle n_{t}^{2}\right\rangle $, i.e. the first and second
moments of $Q_{n}\left(t\right)$ with respect to $n$. Starting from
the expressions
\[
\left\langle \tilde{n}_{s}\right\rangle =\frac{\til{\f}\left(s\right)}{s\left(1-\til{\f}\left(s\right)\right)}\text{ and }\left\langle \tilde{n}_{s}^{2}\right\rangle =\frac{\til{\f}\left(s\right)\left(1+\til{\f}\left(s\right)\right)}{s\left(1-\til{\f}\left(s\right)\right)^{2}},
\]
 and the Laplace-space expansion 
\[
\til{\f}\left(s\right)=\sum_{m=0}^{\infty}c_{m}s^{m}+s^{\b}\sum_{r=0}^{\infty}d_{r}s^{r},
\]
of the main text Eqs. $\left(13\right)$ and $\left(14\right)$, we
first obtain

\begin{equation}
\left\langle \tilde{n}_{s}\right\rangle =s^{-1}\frac{s^{\b}\sum_{r=0}^{\infty}d_{r}s^{r}+\sum_{m=0}^{\infty}c_{m}s^{m}}{1-\left(s^{\b}\sum_{r=0}^{\infty}d_{r}s^{r}+\sum_{m=0}^{\infty}c_{m}s^{m}\right)}.\label{eq:n_s}
\end{equation}
Normalization of $\f\left(\t\right)$ implies that $c_{0}=1$. Consequently,
we approximate $\left\langle \tilde{n}_{s}\right\rangle $ for large
$t$ (i.e. small $s$) as
\begin{equation}
\left\langle \tilde{n}_{s}\right\rangle \approx-c_{1}^{-1}s^{-2}\left[1-\frac{d_{0}}{c_{1}}s^{\b-1}+\left(c_{1}-\frac{c_{2}}{c_{1}}\right)s\right],\label{eq:n_s_1}
\end{equation}
neglecting higher order terms in $s$. Using the expression for the
inverse Laplace-transform 
\begin{equation}
\L^{-1}\left[s^{\g}\right]=\frac{t^{-1-\g}}{\G\left[-\g\right]},\label{eq:inv-Lap}
\end{equation}
where $\G\left[x\right]$ denotes the Euler gamma function, one finds
\begin{equation}
\left\langle n_{t}\right\rangle \asy-\frac{t}{c_{1}}+\frac{d_{0}t^{2-\b}}{\G\left[3-\b\right]c_{1}^{2}}+\frac{c_{2}-c_{1}^{2}}{c_{1}^{2}},\label{eq:n_t}
\end{equation}
as in the main text Eq. $\left(15\right)$. The same is done for the
second moment, for which we find 
\[
\left\langle \tilde{n}{}_{s}^{2}\right\rangle \approx c_{1}^{-2}s^{-3}\left(1+c_{1}s\right)\left(2+c_{1}s\right)
\]
\begin{equation}
\times\left[1-2\left(\frac{d_{0}}{c_{1}}s^{\b-1}+\frac{c_{2}}{c_{1}}s\right)+3\left(\frac{d_{0}}{c_{1}}s^{\b-1}+\frac{c_{2}}{c_{1}}s\right)^{2}\right],\label{eq:n_s_sqr}
\end{equation}
neglecting higher order terms in small $s$. Taking the inverse Laplace
transform then yields 
\[
\left\langle n_{t}^{2}\right\rangle \approx-\frac{4d_{0}}{\G\left[4-\b\right]c_{1}^{3}}t^{3-\b}+\frac{t^{2}}{c_{1}^{2}}
\]
\begin{equation}
+\frac{\left(3c_{1}^{2}-4c_{2}\right)t}{c_{1}^{3}}+\frac{6d_{0}t^{4-2\b}}{\G\left[5-2\b\right]c_{1}^{4}},\label{eq:n_t_sqr}
\end{equation}
as in the main text Eq. $\left(16\right)$ where, again, we neglect
higher order terms in large $t$. With this, we obtain the large $t$
fluctuations $\left\langle \D n_{t}^{2}\right\rangle \equiv\left\langle n_{t}^{2}\right\rangle -\left\langle n_{t}\right\rangle ^{2}$
in the main text Eq. $\left(17\right)$. 

\section{Small-$s$ Expansion of $\tilde{\protect\f}\left(s\right)$}

We here provide the series expansion of $\tilde{\f}\left(s\right)=\int_{0}^{\infty}\dif te^{-st}\f\left(t\right)$
for $\f\left(\t\right)=\b\th\left[\t-1\right]\t^{-\left(1+\b\right)}$
of the main text Eq. $\left(1\right)$. We then explicitly show that
the same general expansion $\til{\f}\left(s\right)=\sum_{m=0}^{\infty}c_{m}s^{m}+s^{\b}\sum_{r=0}^{\infty}d_{r}s^{r}$
which appears in the main text Eq. $\left(14\right)$, similarly applies
for two other choices of walk-time distribution. 

For the particular walk-time distribution $\f\left(\t\right)=\b\th\left[\t-1\right]\t^{-\left(1+\b\right)}$,
the Laplace transform $\til{\f}\left(s\right)$ is given by 
\begin{equation}
\til{\f}\left(s\right)=\b\int_{1}^{\infty}\dif te^{-st}t^{-1-\b}\equiv\b E_{1+\b}\left[s\right],\label{eq:phi_s}
\end{equation}
where $E_{p}\left[z\right]=\int_{1}^{\infty}\dif ye^{-zy}y^{-p}$
is the generalized exponential integral. Using its series expansion
$E_{p}\left[z\right]=\G\left[1-p\right]z^{p-1}-\sum_{m=0}^{\infty}\frac{\left(-z\right)^{m}}{m!\left(1-p+m\right)}$
 \citep{olver2010nist}, the series expansion
\begin{equation}
\til{\f}\left(s\right)=-\G\left[1-\b\right]s^{\b}-\b\sum_{m=0}^{\infty}\frac{\left(-s\right)^{m}}{m!\left(m-\b\right)},\label{eq:phi_s_series}
\end{equation}
sets the coefficients $c_{1}=-\frac{\b}{\b-1}$, $d_{0}=-\G\left[1-\b\right]$
and $c_{2}=-\frac{\b}{2\left(2-\b\right)}$ that appears in the main
text. 

For completeness, we next consider two additional choices of walk-time
distributions, $\c\left(\t\right)=\b\left(1+\t\right)^{-1-\b}$ and
$\z\left(\t\right)=\left(\G\left[\b\right]e^{1/\t}\t^{1+\b}\right)^{-1}$,
which feature a heavy tail $\pro\t^{-1-\b}$ for large $\t$ and $1<\b<2$,
and show that both are consistent with the form $\til{\f}\left(s\right)=\sum_{m=0}^{\infty}c_{m}s^{m}+s^{\b}\sum_{r=0}^{\infty}d_{r}s^{r}$
in the main text Eq. $\left(14\right)$. The Laplace transform of
$\c\left(\t\right)=\b\left(1+\t\right)^{-1-\b}$ is given by $\tilde{\c}\left(s\right)=\b e^{s}E_{1+\b}\left(s\right)$.
Using the series expansions of $E_{1+\b}\left(s\right)$ and $e^{s}$,
one finds $\tilde{\c}\left(s\right)=\b\biggl(s^{\b}\sum_{\el=0}^{\infty}\frac{\G\left[-\b\right]}{\el!}s^{\el}$$-\sum_{m,\el=0}^{\infty}\frac{\left(-1\right)^{m}}{\el!m!\left(m-\b\right)}s^{\el+m}\biggl)$.
By converting the double sum $\sum_{m,\el=0}^{\infty}\frac{\left(-1\right)^{m}}{\el!m!\left(m-\b\right)}s^{\el+m}$
into a single sum, we obtain
\begin{equation}
\tilde{\c}\left(s\right)=\sum_{q=0}^{\infty}\frac{\G\left[1-\b\right]}{\left(q-\b\right)!}s^{q}+s^{\b}\sum_{\el=0}^{\infty}\frac{\b\G\left[-\b\right]}{\el!}s^{\el},\label{eq:chi series}
\end{equation}
which is precisely the form $\til{\f}\left(s\right)=\sum_{m=0}^{\infty}c_{m}s^{m}+s^{\b}\sum_{r=0}^{\infty}d_{r}s^{r}$.
The same happens for the Laplace transform of $\z\left(\t\right)$,
which is given by $\tilde{\z}\left(s\right)=2\G\left[\b\right]^{-1}s^{\b/2}K_{\b}\left(2\sqrt{s}\right)$,
where $K_{\a}\left(x\right)$ is the modified Bessel function of the
second kind. This may be represented by the series
\[
\tilde{\z}\left(s\right)=\sum_{m=0}^{\infty}\frac{\G\left[1-\b\right]}{m!\left(m-\b\right)!}s^{m}
\]
\begin{equation}
+s^{\b}\sum_{\el=0}^{\infty}\frac{-\pi}{\sin\left[\pi\b\right]\G\left[\b\right]\el!\left(\el+\b\right)!}s^{\el},\label{eq:zeta series}
\end{equation}
which, again, perfectly agrees with the form $\til{\f}\left(s\right)=\sum_{m=0}^{\infty}c_{m}s^{m}+s^{\b}\sum_{r=0}^{\infty}d_{r}s^{r}$.

\section{Continuous Time Random Walk }

In this section we analyze the onset of superdiffusion and the walk-number
distribution of the continuous time random walk (CTRW) model. We show
that, even though $\left\langle \d n_{t}^{2}\right\rangle $ does
exhibit a transition at $\b_{c}$ (as in the main text Eq. $\left(17\right)$
for the L\'evy walk), the absence of a coupling of the particle's
position to time prevents a corresponding transition in the onset
of superdiffusion. We denote by $\n_{n}^{CTRW}\left(x,t\right)$ the
density per unit-time of particles making their $n$'th jump at position
$x$ at time $t$ and by $P_{n}^{CTRW}\left(x,t\right)$ the density
of particles which have made $n$ jumps and are currently located
at position $x$ at time $t$. The jump-distance distribution is denoted
by $g\left(\el\right)$ and the waiting-time distribution is denoted
by $\o\left(\s\right)$, with the survival probability given by $\O\left(\s\right)=\int_{\s}^{\infty}\dif\t\o\left(\t\right)$
(i.e. the probability of waiting a time \textit{greater} than $\s$
before the next jump). The equation describing the evolution of $\n_{n}^{CTRW}\left(x,t\right)$
is
\[
\n_{n+1}^{CTRW}\left(x,t\right)=\int_{-\infty}^{\infty}\dif yg\left(y\right)
\]
\begin{equation}
\times\int_{0}^{t}\dif\t\o\left(\t\right)\n_{n}^{CTRW}\left(x-y,t-\t\right),\label{eq:ctrw nu}
\end{equation}
with the initial condition $\n_{1}^{CTRW}\left(x,t\right)=\d\left[x\right]\o\left(t\right)$
ensuring that particles make their first jump (i.e. $n=1$) at the
origin after waiting a time $t$. Similarly, the equation for $P_{n}^{CTRW}\left(x,t\right)$
is
\[
P_{n}^{CTRW}\left(x,t\right)=\int_{-\infty}^{\infty}\dif yg\left(y\right)
\]
\begin{equation}
\times\int_{0}^{t}\dif\t\O\left(\t\right)\n_{n}^{CTRW}\left(x-y,t-\t\right).\label{eq:ctrw P}
\end{equation}
Taking the Fourier-Laplace transform of Eq. (\ref{eq:ctrw nu}), using
its initial condition and solving the $n$ dependence yields
\begin{equation}
\til{\n}_{n}^{CTRW}\left(k,s\right)=\hat{g}\left(k\right)^{n-1}\tilde{\o}\left(s\right)^{n}.\label{eq:ctrw nu Laplace}
\end{equation}
Repeating this for $\tilde{P}_{n}^{CTRW}\left(k,s\right)$ yields
\begin{equation}
\tilde{P}_{n}^{CTRW}\left(k,s\right)=s^{-1}\left(1-\tilde{\o}\left(s\right)\right)\left(\hat{g}\left(k\right)\tilde{\o}\left(s\right)\right)^{n},\label{eq:ctrw P Laplace}
\end{equation}
where we have also used $\tilde{\O}\left(s\right)=s^{-1}\left(1-\tilde{\o}\left(s\right)\right)$.

Let us first study the jump-number distribution $\tilde{Q}_{n}^{CTRW}\left(s\right)$
which describes the number of jumps $n$ completed at time $t$. To
this end, we marginalize $\tilde{P}_{n}^{CTRW}\left(k,s\right)$ over
position by setting $k=0$ and find
\begin{equation}
\tilde{Q}_{n}^{CTRW}\left(s\right)=s^{-1}\left(1-\tilde{\o}\left(s\right)\right)\tilde{\o}\left(s\right)^{n}.\label{eq:ctrw Q}
\end{equation}
This is precisely the same distribution found in the main text Eq.
$\left(11\right)$ for the L\'evy walk model. As such, it entails
the same transition in the pre-asymptotic fluctuations $\left\langle \d n_{t}^{2}\right\rangle $
in the main text Eq. $\left(17\right)$, naively suggesting a corresponding
transition in the onset of superdiffusion, as in the L\'evy walk.
However, as we next demonstrate, the absence of a speed $v$ which
couples the particle's position to time consequently implies that
these fluctuations cannot enter the CTRW propagator. Consequently,
no transition will occur in the onset of superdiffusion in the CTRW
model.

To compute the propagator, we marginalize over the number of jumps
$n$ and obtain the known CTRW propagator \citep{zaburdaev2015levy}
\begin{equation}
\tilde{P}^{CTRW}\left(k,s\right)=\frac{1-\tilde{\psi}\left(s\right)}{s\left(1-\hat{g}\left(k\right)\tilde{\psi}\left(s\right)\right)}.\label{eq:ctrw prop}
\end{equation}
Let us choose the specific waiting-time and jump-length distributions
\begin{equation}
\begin{cases}
\o\left(\t\right)=\b\th\left[\t-1\right]\t^{-1-\b}\\
g\left(\el\right)=\frac{\g}{2}\th\left[\left|\el\right|-1\right]\left|\el\right|^{-1-\g}
\end{cases},\label{eq:ctrw distributions}
\end{equation}
whose respective Laplace and Fourier transforms are 
\begin{equation}
\tilde{\o}\left(s\right)=-\G\left[1-\b\right]s^{\b}-\b\sum_{m=0}^{\infty}\frac{\left(-s\right)^{m}}{m!\left(m-\b\right)},\label{eq:ctrw transformed omega}
\end{equation}
and 
\[
\hat{g}\left(k\right)=\hg 12\left[-\frac{\gamma}{2};\frac{1}{2},\frac{2-\gamma}{2};-\frac{k^{2}}{4}\right]
\]
\begin{equation}
-\cos\left[\frac{\pi\gamma}{2}\right]\Gamma\left[1-\gamma\right]\left|k\right|^{\gamma},\label{eq:ctrw transformed g}
\end{equation}
where $\b>1$ and $1<\g<2$ correspond to the superdiffusive regime
and $\hg pq\left[\left\{ a_{i}\right\} _{i=1}^{p};\left\{ b_{j}\right\} _{j=1}^{q};z\right]$
is the generalized hypergeometric function. For small $s$ and $\left|k\right|$,
these become
\begin{equation}
\begin{cases}
\tilde{\o}\left(s\right)\approx1-\frac{\b}{\b-1}s-\G\left[1-\b\right]s^{\b}-\frac{\b}{2\left(2-\b\right)}s^{2}\\
\hat{g}\left(k\right)\approx1-\cos\left[\frac{\pi\g}{2}\right]\G\left[1-\g\right]\left|k\right|^{\g}+\frac{\g}{2\left(2-\g\right)}k^{2}
\end{cases}.\label{eq:ctrw distributions series}
\end{equation}
Taking first the limit of large $t$ and then the limit of larges
distances yields
\begin{equation}
\tilde{P}^{CTRW}\left(k,s\right)\asy\frac{\b}{\b-1}\frac{1}{1-\hat{g}\left(k\right)+\frac{\b}{\b-1}\hat{g}\left(k\right)s},\label{eq:ctrw prop asy}
\end{equation}
whose Inverse Laplace transform at small $\left|k\right|$ is 
\begin{equation}
\tilde{P}^{CTRW}\left(k,t\right)\asy\hat{g}\left(k\right)^{-1}e^{-\frac{\b-1}{\b}t\frac{1-\hat{g}\left(k\right)}{\hat{g}\left(k\right)}}.\label{eq:ctrw prop inv Lap}
\end{equation}
For small $\left|k\right|$ we find
\begin{equation}
\hat{P}\left(k,t\right)\asy e^{-\frac{\left(\b-1\right)t}{\b}\left(\cos\left[\frac{\pi\g}{2}\right]\G\left[1-\g\right]\left|k\right|^{\g}-\frac{\g k^{2}}{2\left(2-\g\right)}+\ord{\left|k\right|^{2+\g}}\right)},\label{eq:ctrw prop fin}
\end{equation}
with no transition in the onset of superdiffusion.

\section{L\'evy Flight}

We here analyze the onset of superdiffusion in the L\'evy flight
model and demonstrate the absence of transition in the onset of superdiffusion.
Denoting the $P_{n}^{LF}\left(x\right)$ density of fliers at position
$x$ after $n$ steps, the model's evolution is described by
\begin{equation}
P_{n+1}^{LF}\left(x\right)=\int_{-\infty}^{\infty}\dif y\f\left(y\right)P_{n}^{LF}\left(x-y\right).\label{eq:LF P eqn}
\end{equation}
Following a Fourier transform, we obtain
\begin{equation}
\hat{P}_{n}\left(k\right)=\hat{\f}\left(k\right)^{n}.\label{eq:LF P k space}
\end{equation}
Let us choose the flight length distribution $\f\left(\el\right)=\frac{\b}{2}\th\left[\left|\el\right|-1\right]\left|\el\right|^{-1-\b}$,
whose Fourier transform is given by 
\[
\hat{\f}\left(k\right)=\hg 12\left[-\frac{\b}{2};\frac{1}{2},\frac{2-\b}{2};-\frac{k^{2}}{4}\right]
\]
\begin{equation}
-\cos\left[\frac{\pi\b}{2}\right]\Gamma\left[1-\b\right]\left|k\right|^{\b}.\label{eq:LF phi transformed}
\end{equation}
For small $\left|k\right|$, this becomes
\[
\hat{\f}\left(k\right)=1-\cos\left[\frac{\pi\b}{2}\right]\Gamma\left[1-\b\right]\left|k\right|^{\b}
\]
\begin{equation}
+\frac{\b}{2\left(2-\b\right)}k^{2}+\ord{k^{4}}.\label{eq:LF phi series}
\end{equation}
Thus, for small $\left|k\right|$, $\hat{P}_{n}\left(k\right)$ is
well approximated by
\begin{equation}
\hat{P}_{n}\left(k\right)\asy e^{-n\left(D_{0}\left|k\right|^{\b}-D_{1}k^{2}+\ord{k^{4}}\right)},\label{eq:LF asy prop}
\end{equation}
where
\begin{equation}
\begin{cases}
D_{0}=\cos\left[\frac{\pi\b}{2}\right]\Gamma\left[1-\b\right]\left|k\right|^{\b}\\
D_{1}=\frac{\b}{2\left(2-\b\right)}
\end{cases}.\label{eq:LF coefficients}
\end{equation}
Again, no transition is found in the onset of superdiffusion.

\end{document}